\documentclass[preprint,aps,amssymb,showpacs,superscriptaddress,nofootinbib]{revtex4}
\usepackage{graphicx}
\newcommand{\del}{\partial}
\newcommand{\f}{\frac}
\newcommand{\Case}[2]{{\textstyle \frac{#1}{#2}}}

\begin{document}
\preprint{IMSc/2007/12/15}

\title{Loop Quantization of Polarized Gowdy Model on $T^3$: Classical
Theory}

\author{Kinjal Banerjee}
\email{kinjal@imsc.res.in}
\affiliation{The Institute of Mathematical Sciences\\
CIT Campus, Chennai-600 113, INDIA.}

\author{Ghanashyam Date}
\email{shyam@imsc.res.in}
\affiliation{The Institute of Mathematical Sciences\\
CIT Campus, Chennai-600 113, INDIA.}

\begin{abstract} 
The vacuum Gowdy models provide much studied, non-trivial
midi-superspace examples. Various technical issues within Loop Quantum
Gravity can be studied in these models as well as one can hope to
understand singularities and their resolution in the loop quantization.
The first step in this program is to reformulate the model in real
connection variables in a manner that is amenable to loop quantization.
We begin with the unpolarized model and carry out a consistent reduction
to the polarized case. Carrying out complete gauge fixing, the known
solutions are recovered.
\end{abstract}


\maketitle

\section{Introduction}

Gowdy space-times \cite{Gowdy} are solutions of vacuum Einstein equation
admitting two commuting, space-like isometries and having closed spatial
hypersurfaces.  Such models have essentially one of the three topologies
for the spatial slices: $T^3, S^3, S^1 \times S^2$. The case of $T^3$ is
the simplest of these. A further simplification is possible. One can
restrict to the so-called {\em polarized} models in which the two
Killing vectors are orthogonal. This simplest case of polarized, $T^3$
vacuum Gowdy model is the focus of this series of works. In this case,
the complete set of exact solutions is known \cite{Moncrief} which
generically have {\em initial singularity}. But there is also an
infinite sub-family of solutions for which all curvature invariants are
finite.  The approach to classical singularity is well studied and is
known to follow a special case of the BKL scenario known as
asymptotically velocity term dominated near singularity (AVTDS)
\cite{IsenbergMoncrief}. At late times, the model is
known to be asymptotically homogeneous \cite{Berger}.

These models have been analysed in the canonical framework in both
metric variables as well as in terms of the complex Ashtekar variables.
The first attempts of quantization, were carried out in ADM variables in
\cite{ADMQuantization}.  Another approach which has been more successful
was based on an interesting property of the model. After a suitable
(partial) gauge fixing, these models can be described by (modulo a
remaining global constraint) a ``point particle'' degree of freedom and
by a scalar field $\phi$ which is subject to the same equations of
motion as a massless, rotationally symmetric, free scalar field
propagating in a fictitious two dimensional expanding torus.  This
equivalence was used in the quantization carried out in \cite{Pierri}.
Subsequent analysis has been carried out in a large number of works some
of which are listed in \cite{MenamaruganTorre}. However in these
quantizations, the evolution turned out to be non-unitary and in
\cite{Corichi} a new parametrization was introduced which implemented
unitary evolution in quantum theory.

Canonical description of {\em unpolarised} Gowdy $T^3$ model in terms of
the complex Ashtekar variables has been given in \cite{HusainSmolin}.  A
complete set of Dirac observables is also known \cite{Husain}.  The
canonical quantization of this model was carried out in
\cite{Menamarugan} and the physical Hilbert space was obtained.
Although quantization has been carried out, the difficult issue of
singularities has not been fully addressed (see however
\cite{HusainSingularity} for preliminary attempts). 

In this series of works, we aim to carry out {\em loop quantization} of
the polarized, $T^3$ Gowdy model, obtained via symmetry reduction in
terms of real connection variables.  In this paper, we report the first
step of recasting the Gowdy model in the (real) connection formulation,
including the restriction to the polarized model. In subsequent papers
the quantization will be carried out and issue of singularities will be
addressed.

In section \ref{2} we discuss, in brief, Gowdy models and then restrict
attention to the polarised $T^3$ case in metric variables.  The form of
the metric and the space-time solutions are discussed.  In section
\ref{3} the unpolarised model is described in terms of real Ashtekar
variables and a {\em consistent} reduction is carried out to obtain the
polarised model. Here consistency refers to preservation of the diagonal
form of the metric under the Hamiltonian evolution.  In section \ref{4},
we explain a gauge fixing leading to recovery of the standard form of
the solutions.  Section \ref{5} summarizes the results and includes
preliminary comments.

\section{Polarised Gowdy $T^3$ Model in Metric Variables} \label{2}

Gowdy space-times \cite{Gowdy} are globally hyperbolic solutions of the
vacuum Einstein's equations which are isometric under the action of the
Abelian group $T^2$ which acts on the spatial slices assumed to be
closed. This means that there are two independent commuting spatial
Killing vectors.  This condition along with the Einstein's equations
restricts the allowable choices of spatial topology to be only $T^3$,
$S^3$ and $S^2 \times S^1$ (or certain manifolds with one of these as
cover).  If, in addition we make an additional assumption that the
Killing vector fields which generate the $T^2$ isometry can be chosen to
be mutually orthogonal everywhere, we get the so-called {\em polarized
Gowdy model}.  The metric, is then diagonal and can be written as
\cite{Moncrief}:
\begin{eqnarray}
\text{d}s^2=e^{2a}(-\text{d}T^2 + \text{d}\theta^2) + T
(e^{2W}\text{d}x^2 + e^{-2W}\text{d}y^2)  \label{metricmetric}
\end{eqnarray} 
where ${\del}/{\del x}$ and ${\del}/{\del y}$ are the two Killing
vectors and  $a$ and $W$ are functions of $T$ and periodic functions of
$\theta$.

The solutions for W can be obtained from the second order differential
equation:
\begin{eqnarray}
\f{\del^2 W}{\del T^2} + \f{1}{T}\f{\del W}{\del T} - \f{\del^2 W}{\del
\theta^2} = 0 \label{Weqn}
\end{eqnarray}
Given a solution, $W(T, \theta)$, the function $a(T, \theta)$ has to
satisfy:
\begin{eqnarray}
\f{\del a}{\del \theta} &=& 2T \f{\del W}{\del T}~ \f{\del W}{\del
\theta}  \nonumber \\
\f{\del a}{\del T} &=& - \f{1}{4T} + T \left[ \left(\f{\del W}{\del
T}\right)^2 + \left(\f{\del W}{\del \theta}\right)^2 \right]
\label{aeqn}
\end{eqnarray}
The $W$ equation (\ref{Weqn}) encodes the dynamics while the
(\ref{aeqn}) encodes the constraints.  Incidentally, this makes the
initial value problem and the problem of preservation of constraints in
numerical relativity trivial for this model.  The initial values of the
dynamical variable $W$ can be freely specified and given a $W$ the
constraint $a$ can be trivially determined \cite{BergerGarfinkle}. The
requirement of $a$ being a periodic function of $\theta$ imposes a
condition of the solutions of the $W$ equation, namely, $\int d\theta
\partial_T W \partial_{\theta} W = 0$. The general solution given below,
does satisfy this condition \cite{Moncrief}. 

The general solution to (\ref{Weqn}) is given by
\begin{eqnarray}
W = \alpha +\beta \mbox{ln}~ T + \sum^{\infty}_{n=1} \bigg[a_n J_0 (nT)
\mbox{sin} (n \theta + \gamma_n) + b_n N_0 (nT) \mbox{sin} (n \theta
+\delta_n )\bigg] \label{Wsolution} 
\end{eqnarray}
where $\alpha$, $\beta$, $a_n$, $b_n$, $\gamma_n$ and $\delta_n$ are
real constants and $J_0$ and $N_0$ are regular and irregular Bessel
functions of the zeroth order.  The special case of {\em homogeneous}
model is given by $\beta = \f{1}{2}$ and $a_n~=0~=b_n$ and corresponds
to the flat Kasner solution described as:
\begin{eqnarray}
\text{d}s^2 = -\text{d}T^2 + \text{d}\theta^2 + T^2 \text{d}x^2 +
\text{d}y^2
\end{eqnarray}

It can be shown that the curvature invariant $C\equiv R_{abcd} R^{abcd}$
blows up almost everywhere as $T~\rightarrow~0_+$.  The solutions are
therefore generically singular. However for the special choice, 
\begin{eqnarray}
b_n = 0 ~\mbox{,}\hspace{2em} \beta = \f{1}{2}
\end{eqnarray}
the curvature invariant remains bounded and all components of $R_{abcd}$
have finite limit as $T~\rightarrow~0_+$.  It can also be shown that
these nonsingular solutions are analytically extendible \cite{Moncrief}
but are causally ill-behaved (have closed time-like curves) in the
extended portion. Thus there exists a infinite number of nonsingular
solutions which however form a set of measure zero in the space of
solutions. Curvature unboundedness is the generic behaviour.

\section{Polarised Gowdy $T^3$ Model in real Ashtekar Variables} \label{3}

\subsection{Unpolarised Case}

We begin with the connection formulation in the notation of
\cite{AshtekarLewandowski}, with $P^i_a = E^i_a/(\kappa\gamma)$
substituted. The classical symmetry reduction to the unpolarized Gowdy
$T^3$ model in terms of the complex connection has been studied by
\cite{HusainSmolin,Menamarugan}.  Translated into real variables, the
symmetry reduction is achieved by setting to zero the following
components of the densitized triad and connection \cite{Spherical1}: 
\begin{equation}
E^{\theta}_I ~=~ 0 ~=~ E^{\rho}_3 ~~,~~  A_{\theta}^I ~=~ 0 ~=~
A_{\rho}^3   ~~~;~~~ \rho ~=~ x,\ y ~~;~~ I ~=~ 1,\ 2\ .
\end{equation}

In these variables, the Gauss, the diffeomorphism and the Hamiltonian
constraints \cite{AshtekarLewandowski}, are given by ($\kappa := 8 \pi
G_{\mathrm {Newton}}$, $G := G_3$, $C := C_{\theta}$):
\begin{eqnarray}
G &=& \f{1}{\kappa\gamma}\left[ \del_{\theta} E^{\theta}_3
+\epsilon_J^{\ K} A_{\rho}^J E^{\rho}_K \right]~ ;
\hspace{5.0cm}\epsilon_J^{\ K} ~ := ~ \epsilon_{3 J}^{~\
K}\label{gauss1}\\ 
C &=& \f{1}{\kappa\gamma}\left[(\del_\theta  A_{\rho}^I) E^{\rho}_I +
\epsilon_J^{\ K} A_{\rho}^J E^{\rho}_K A_{\theta}^3 - \kappa\gamma
A_\theta^3 G_3\right]~ ;  \label{diffeo1}\\ 
H &=& \f{1}{2\kappa} \f{1}{\sqrt{|\mbox{det} E|}}\left[~ 2 A_{\theta}^3
E^{\theta}_3 A_{\rho}^J E^{\rho}_J + A_{\rho}^J E^{\rho}_J A_{\sigma}^K
E^{\sigma}_K - A_{\rho}^K E^{\rho}_J A_{\sigma}^J E^{\sigma}_K - 2
\epsilon_J^{\ K} (\del_{\theta} A_{\rho}^J) E^{\rho}_K
E^{\theta}_3\right.  \nonumber\\ 
& & \hspace{2.6cm} - \left. (1+\gamma^2) \left( 2 K_{\theta}^3
E^{\theta}_3 K_{\rho}^J E^{\rho}_J + K_{\rho}^J E^{\rho}_J K_{\sigma}^K
E^{\sigma}_K - K_{\rho}^K E^{\rho}_J K_{\sigma}^J
E^{\sigma}_K\right)\right]~ .  \label{hamiltonian1}
\end{eqnarray}
In the above, $K^i_a$ are the components of the extrinsic curvature
which are related to the gravitational connection $A^i_a$ and the
torsion-free spin connection, $\Gamma^i_a$, as: $K^i_a =
\gamma^{-1}(A^i_a - \Gamma^i_a)$ and $\gamma$ is the Barbero-Immirzi
parameter. The spin-connection is defined in eqn.
(\ref{SpinConnection}).

Since none of the quantities depend on $x$ or $y$ we can integrate over
the $T^2$ and write the symplectic structure and the total Hamiltonian
as:
\begin{eqnarray}
\Omega & = & \f{4\pi^2}{\kappa\gamma}\int\text{d}\theta\left(
\text{d}A_{\theta}^3 \wedge \text{d}E^{\theta}_3 + \text{d}A_{\rho}^I
\wedge \text{d}E^{\rho}_I \right) \label{symplectic1} \\
H_{\mathrm{tot}} & = & 4 \pi^2 \int d\theta \left\{\lambda^3 G +
N^{\theta} C + N H\right\} \label{TotalHamiltonian}
\end{eqnarray}

Under the $\theta$ coordinate transformation $E^{\theta}_3$ transforms
as a scalar, $E^{\rho}_I$'s transform as scalar densities of weight 1,
$A_{\theta}^3$ transforms as a scalar density of weight 1 and
$A_{\rho}^I$'s transform as scalars \footnote{In one dimension, under
orientation preserving coordinate transformations, a tensor density of
contravariant rank $p$, covariant rank $q$ and weight $w$, can be
thought of as a scalar density of weight $= w + q - p$.}.  

For each $\rho$, the $A_{\rho}^I$ and $E_I^{\rho}$, rotate among
themselves under the $U(1)$ gauge transformations generated by the Gauss
constraint.  It is however possible to choose variables which are gauge
invariant and will turn out to be more suitable for loop quantization
(see section \ref{5}).  These are introduced through the following
definitions:
\begin{eqnarray} 
E^{{x}}_1 &=& E^x \mbox{cos} \beta  \hspace{4em} ~ ~ \text{ ; }
\hspace{0.6cm} E^{{x}}_2 = E^x \mbox{sin} \beta \label{AngleOne}\\
E^{{y}}_1 &=& - E^y \mbox{sin} \bar\beta \hspace{4em} \text{ ; }
\hspace{0.6cm} E^{{y}}_2 = E^y \mbox{cos} \bar\beta \label{AngleTwo}\\
A_{{x}}^1 &=& A_{{x}}  \mbox{cos} ( \alpha + \beta) \hspace{0.6cm} ~ ~
\text{ ; } \hspace{0.6cm} A_{{x}}^2 = A_{{x}}  \mbox{sin} ( \alpha +
\beta) \label{AngleThree}\\
A_{{y}}^1 &=& - A_{{y}}  \mbox{sin} ( \bar\alpha + \bar\beta)
\hspace{0.6cm}  \text{ ; } \hspace{0.6cm} A_{{y}}^2 = A_{{y}} \mbox{cos}
(\bar\alpha +\bar\beta\label	{AngleFour})
\end{eqnarray}
The angles for the connection components are introduced in a particular
fashion for later convenience.

The radial coordinates, $E^x, E^y, A_x, A_y$, are gauge invariant and
always strictly positive (vanishing radial coordinates correspond to
trivial symmetry orbit which is ignored).  

In terms of these variables, the symplectic structure
(\ref{symplectic1}) gets expressed as:
\begin{eqnarray}
\Omega=\f{4\pi^2}{\kappa\gamma}\int\text{d}\theta\left[
\text{d}A^3_{\theta} \wedge \text{d}E_3^{\theta} +  \text{d} X \wedge
\text{d} E^{x} + \text{d} Y \wedge \text{d} E^{y} + \text{d} \beta
\wedge \text{d} P^{\beta} + \text{d} \bar\beta \wedge \text{d} \bar
P^{\beta}\right] \label{symplectic2}
\end{eqnarray}
where:
\begin{eqnarray}
X &:=& A_x \mbox{cos} ( \alpha ) \hspace{2em} ~ ~ ~ \text{ ; }
\hspace{1em} Y := A_y \mbox{cos} (\bar\alpha ) \\
P^{\beta} &:=& - E^x A_x \mbox{sin} ( \alpha ) \hspace{1em}  \text{ ; }
\hspace{1em} \bar P^{\beta} :=  - E^y A_y \mbox{sin} ( \bar\alpha ) 
\end{eqnarray}
The gauge transformations generated by the Gauss constraint shift
$\beta, \bar{\beta}$ rendering $\alpha$ and $\bar\alpha$ gauge
invariant. {\em From now on we will absorb the $4\pi^2$ and use $\kappa'
:= \Case{\kappa}{4\pi^2} = \Case{2G_{\mathrm{Newton}}}{\pi}$}.

It is convenient to make a further canonical transformation:
\begin{eqnarray}
\xi &=& \beta - \bar\beta \hspace{1.0cm} \text{ ; } \hspace{1em} \eta
= \beta + \bar\beta \label{CanOne}\\
P^{\xi} &=& \f{P^{\beta} - \bar P^{\beta}}{2}  \hspace{1em} \text{ ; }
\hspace{1em} P^{\eta} = \f{P^{\beta} + \bar P^{\beta}}{2} \label{CanTwo}
\end{eqnarray}

In terms of these variables the Gauss and the diffeomorphism constraints can be
written as:
\begin{eqnarray}
G &=& \f{1}{\kappa\gamma}\left[ \del_{\theta} E_3^{\theta} + 2
P^{\eta}\right] \label{gauss2} \\
C &=& \f{1}{\kappa\gamma}\left[ (\del_{\theta}X)E^x + (\del_{\theta}Y)E^y -
(\del_{\theta}E_3^\theta)A^3_\theta + (\del_{\theta}\eta)P^{\eta} +
(\del_{\theta}\xi)P^{\xi} \right] \label{vector2} 
\end{eqnarray}
The Hamiltonian constraint is complicated but after  putting $K^a_i =
(A^a_i-\Gamma^a_i)/\gamma$, substituting the explicit expressions of
$\Gamma^a_i$, and further simplification, turns out to be:
\begin{eqnarray}
H = &-&\f{\gamma^{-2}}{2\kappa} \f{1}{\sqrt{E}}\left[
E_3^{\theta}\bigg\{(X E^x + Y E^y)\del_\theta \eta + (X E^x - Y
E^y)\del_\theta \xi -2 P^{\xi} \del_\theta\left(\mbox{ln} \f
{E^y}{E^x}\right)\right. \nonumber \\
&+& \left. 2P^{\eta} ( \del_\theta \mathrm{ln}E_3^{\theta} + (\mbox{tan}
\xi) ~ \del_{\theta} \xi) \bigg\}  + 2\bigg\{ (\mbox{cos}^2\xi)\left(X
E^x  Y E^y + ({P^{\eta}})^2 - ({P^{\xi}})^2 \right)\right.\nonumber \\
&+& \left. (X E^x + Y E^y)E_3^{\theta} A^3_{\theta} \bigg\} + (\mbox{sin}2
\xi) \bigg\{ (X E^x + Y E^y)P^{\xi} -(X E^x - Y E^y)P^{\eta}  \bigg\}
\right. \nonumber \\
&+& \left. \left(\f{1+\gamma^2}{2}\right) \left\{
(\del_{\theta}E_3^\theta)^2 - \left(\f{E_3^\theta \del_\theta \xi
}{\mbox{cos}\xi)}\right)^2 -
\left(\f{E_3^\theta\del_\theta(\mbox{ln}(E^y/E^x))}{(\mbox{cos}\xi)}\right)^2\right\}\right]
\nonumber \\
&-& \f{1}{2\kappa}\del_\theta\left(\f{4 E_3^{\theta}
P^{\eta}}{\sqrt{E}}\right) \label{hamiltonian2}
\end{eqnarray}
where $E=|E_3^\theta E^x E^y (\mbox{cos}\xi)|$.

Under the action of the diffeomorphism constraint $X$, $Y$,
$E_3^\theta$, $\eta$ and $\xi$ transforms as scalars while $E^x$, $E^y$,
$A^3_\theta$, $P^{\eta}$ and $P^{\xi}$ transform as scalar densities of
weight 1. 

This completes the description of the {\em unpolarised} Gowdy $T^3$
Model in the variables we have defined. The number of canonical field
variables is 10 while there is a 3-fold infinity of first class
constraints. There are therefore 2 field degrees of freedom.  We now
need to impose two second class constraints such that the number of
field degrees of freedom are reduced from two to one (as it should be in
the polarized case). 

\subsection{Reduction to Polarized model}

The spatial 3-metric, $g_{ab} := e_a^i e_b^j \delta_{ij}$, with the
co-triad $e_a^i$ defined through $e_a^i E^a_j := \delta^i_j \sqrt{E},
\ e_a^i E^b_i := \delta^b_a \sqrt{E}\ $, is given by, 
\begin{eqnarray}
\mbox{d}s^2= \mbox{cos}\xi \f{E^x E^y}{E_3^\theta} \mbox{d}\theta^2 
+ \f{E_3^\theta}{\mathrm{cos}\xi} \f{E^y}{E^x}\mbox{d}x^2 
+ \f{E_3^\theta}{\mathrm{cos}\xi} \f{E^x}{E^y}\mbox{d}y^2 
- 2 \f{E_3^\theta}{\mathrm{cos}\xi} \mathrm{sin}\xi \ \mbox{d}x \mbox{d}y
\label{connectionmetric}
\end{eqnarray}

For the Killing vectors  ${\del}/{\del x}$ and ${\del}/{\del y}$ to be
orthogonal to each other, the $\mbox{d}x\mbox{d}y$ term in the metric
should be zero.  This implies that the polarization condition is
implemented by restricting to $\xi = 0$ sub-manifold of the phase space
of the unpolarized model. For getting a non-degenerate symplectic
structure, one needs to have one more condition. This condition should
be chosen {\em consistently} in the following sense. 

We expect the two conditions to reduce a field degree of freedom. This
can be viewed in two equivalent ways. The condition $\xi = 0$ makes the
metric diagonal and this property should be preserved under evolution
(i.e. the extrinsic curvature should also be diagonal).  Alternatively,
the unpolarized model is a constrained system and we want to impose two
conditions such that one {\em physical} (field) degree of freedom is
reduced. The extra conditions to be imposed should therefore be {\em
first class} with respect to the constraints of the unpolarized model
i.e. should weakly Poisson-commute with them. 

Indeed, this can be done systematically by viewing $\xi = 0$ as a new
constraint\footnote{The choice $\xi = 0$ also requires $E^{\theta}_3 >
0$ for the spatial metric to have signature ($+, +, +$).  The choice
$\xi = \pi$ would require $E^{\theta}_3 < 0$. From now on $E^{\theta}_3
> 0$ will be assumed.} and demanding its preservation under the
evolution generated by the total Hamiltonian.  Since $\xi = 0$ weakly
Poisson commutes with the Gauss and the diffeomorphism constraints, only
the Poisson bracket with the Hamiltonian constraint is needed.
\begin{equation}
\xi(\theta) \approx 0 ~ ~ ~ , ~ ~ ~ \{\xi(\theta), \int d \theta'
N(\theta')H(\theta')\} \approx 0 ~. ~ \label{Constarint1}
\end{equation}
It follows that, 
\begin{eqnarray}
\dot\xi(\theta) \approx 0 ~ ~\Rightarrow ~ ~\chi(\theta) := 2 P^{\xi} +
E_3^\theta \del_\theta(\mbox{ln}E^y/E^x) ~ \approx 0 \label{Constraint2}
\end{eqnarray}
The Poisson Bracket of $\chi$ with the Hamiltonian turns out to be zero
on the constraint surface i.e. $\dot\chi \approx \chi ~ \approx 0$.
Thus, the reduction to the Polarized model is obtained by imposing the
two {\em polarization constraints} 
\begin{equation}
\xi \approx 0  \hspace{1em} \text{ ; } \hspace{1em} \chi \approx 0
\hspace{1em} \text{ ; } \hspace{1em} \{\xi(\theta), \chi(\theta')\} = 2
\kappa\gamma\delta(\theta - \theta') 
\end{equation}

{\em Remark:} To see that the $\chi \approx 0$ condition follows from
preservation of $g_{xy} = 0$, note that, in the metric formulation, for
the present case, it implies that $\dot{g}_{xy} \sim K_{xy} = K^i_x
e^i_y (= e^i_x K^i_y) = 0$. Using the definition $K^i_a =
\gamma^{-1}(A^i_a - \Gamma^i_a)$ and the expressions given in the
appendix, one can check directly that $K_{xy} = 0 \Leftrightarrow \chi =
0$. Note that this is {\em not} equivalent to requiring orthogonality of
components of the connection, $A^i_x A^i_y = 0$ which would imply
$\alpha = \bar{\alpha}$ (see eqns.  (\ref{AngleThree} and
\ref{AngleFour}). This condition, mentioned in the literature
\cite{Spherical1,Neville}, is very different from the $\chi \approx 0$
condition and is not preserved under evolution.

It follows from (\ref{gauss2}) that $\{\chi,G\} = 0$ And using
(\ref{vector2}), one can see that :
\begin{equation}
\Bigg\{\xi,\int N^\theta C_{\theta} \Bigg\} ~ = ~  N^\theta
\del_\theta\xi \approx 0 ~ ~ , ~ ~ 
\Bigg\{\chi,\int N^\theta C_{\theta} \Bigg\} ~ = ~  \del_\theta(N^\theta
\chi) \approx 0 \ .
\end{equation}

We can solve the polarization constraints strongly and use Dirac
brackets. Symbolically,
\begin{eqnarray}
\{f,g\}^{\star} = \{ f,g\} -
\{f,\xi\}\odot\{\xi,\chi\}^{-1}\odot\{\chi,g\} -
\{f,\chi\}\odot\{\chi,\xi\}^{-1}\odot\{\xi,g\} \nonumber 
\end{eqnarray}
Here $\odot$ denotes appropriate integrations since we have field
degrees of freedom.

Since the polarization constraints weakly commute with all the other
constraints, the constraint algebra in terms of Dirac brackets is same
as that in terms of the Poisson brackets and thus remains unaffected.
Furthermore, equations of motions for all the variables other than $\xi,
P_{\xi}$ also remain unaffected. We can thus set the polarization
constraints strongly equal to zero in all the expressions and continue
to use the original Poisson brackets.

The expressions of the constraints simplify greatly and in particular
the Hamiltonian constraint simplifies to,
\begin{eqnarray}
H & = & - \f{\gamma^{-2}}{2\kappa} \f{1}{\sqrt{E}}\left[ ~
\f{(\kappa\gamma G)^2}{2} + (X E^x + Y
E^y)E_3^{\theta}\partial_{\theta}\eta + 2\bigg\{X E^xY E^y +(X E^x + Y
E^y)E_3^{\theta} A^3_{\theta}\bigg\} \right. \nonumber \\ 
& & \hspace{1.7cm} \left. + \f{\gamma^2}{2}\bigg\{
(\del_{\theta}E_3^\theta)^2
-\left(E_3^\theta\del_\theta\mathrm{ln}(E^y/E^x)\right)^2 \bigg\}
\right] + \f{1}{2\kappa}\del_\theta\left\{\f{2 E_3^\theta
\left(\del_\theta E_3^\theta - \kappa\gamma G\right) }{\sqrt{E}}\right\}
\label{finalhamiltonian}
\end{eqnarray}
where, we have also eliminated $P_{\eta}$ in terms of the Gauss
constraint using  $2P^{\eta} = (\kappa\gamma G - \del_{\theta}
E_3^{\theta})$ and $E=|E_3^\theta| E^x E^y$.

Noting that $\eta$ is translated under a gauge transformation, we can
set it to any constant and fix the gauge transformation freedom.
Explicitly, imposing $\eta \approx 0$ as a constraint, we can fix the
$\lambda^3$ from preservation of this gauge fixing condition. Once again
we can use Dirac brackets with respect to the Gauss constraint and the
$\eta \approx 0$ constraint and impose these constraints strongly.  With
this done, the first two terms and the $G$ dependent piece in the last
term in the Hamiltonian, drop out and so do the degrees of freedom
$\eta, P_{\eta}$. We are left with six canonical degrees of freedom and
the two first class constraints, leaving one field degree of freedom.
Thus our final variables and constraints for the {\em polarized} Gowdy
model are (absorbing away the Immirzi parameter): 
\begin{eqnarray}
\kappa := \f{8\pi G_{\mathrm{Newton}}}{4\pi^2}~ , ~ {\cal E}:=
E_3^{\theta}~ , ~ {\cal A} := \gamma^{-1} A^3_{\theta}~ , ~ K_x :=
\gamma^{-1} X~ , ~ K_y := \gamma^{-1} Y \label{NewDefs}\\ 
\{K_x(\theta), E^x(\theta')\} = \kappa \delta(\theta - \theta') ~ ~,~ ~
(\mathrm{and ~ similarly ~ for ~} (K_y, E^y), ({\cal A}, {\cal E})
~\mathrm{ pairs});
\end{eqnarray}
\begin{eqnarray} 
C & = & \f{1}{\kappa}\left[ (\del_{\theta}K_x)E^x +
(\del_{\theta}K_y)E^y - (\del_{\theta}\cal{E})\cal{A} \right]
\label{FinalVector} \\
H & = & \f{1}{\kappa} \left[ - \f{1}{\sqrt{E}} \bigg\{(K_x E^x K_y E^y)
+ (K_x E^x +  K_y E^y)\cal{E} \cal{A}\bigg\} \right. \nonumber\\ 
& & \hspace{0.6cm} \left. - \f{1}{4 \sqrt{E}}\bigg\{
(\del_{\theta}{\cal{E}})^2
-\left({\cal{E}}\del_\theta(\mbox{ln}(E^y/E^x))\right)^2 \bigg\} +
\del_\theta\left(\f{\cal{E}\del_\theta \cal{E} }{\sqrt{E}}\right)
\right] \label{FinalHam}
\end{eqnarray}

The constraint algebra among the $C[N^{\theta}], H[N]$ is:
\begin{eqnarray}
\left\{\ C[N^{\theta}]\ , C[M^{\theta}]\ \right\} & = & C\left[ \
N^{\theta}\partial_{\theta}M^{\theta} -
M^{\theta}\partial_{\theta}N^{\theta}\ \right] \\
\left\{\ C[N^{\theta}]\ , H[N]\ \right\} & = &
H[\ N^{\theta}\partial_{\theta} N\ ] \\
\left\{\ H[M]\ , H[N]\ \right\} & = & C\left[ \ (M\partial_{\theta}N -
N\partial_{\theta}M){\cal E}^2 E^{-1}\ \right] 
\end{eqnarray}
Since each term in the Hamiltonian constraint is a scalar density of
weight +1 and each term in the diffeomorphism constraint is of density
weight +2, the first two brackets are easily verified. The last one also
follows with a bit longer computation.  We have thus verified the
constraint algebra of polarized model showing the consistency of the
reduction procedure. 

\section{Space-time construction} \label{4}

The next task is to find the set of gauge {\em inequivalent} solutions
of the Hamilton's equations of motion, satisfying the two sets of
constraints and obtain the space-time interpretation. The total
Hamiltonian being a constraint, the Lagrange multipliers -- the lapse
function and the shift vector -- also enter in the Hamilton's equations
of motion. These need to be either {\em prescribed} or deduced via a
gauge-fixing procedure.  Once this is done, one can obtain the solution
curves in the phase space with ``initial points'' lying on the
constrained surface. The space-time metric, solving Einstein equation is
then given by,
\begin{equation} \label{SpaceTimeMetric}
ds^2 ~ = ~ - N^2(t, x^i) dt^2 + g_{ij}(t, x^i)\left(dx^i - N^i(t,
x^i)dt\right)\left(dx^j - N^j(t, x^i)dt\right) 
\end{equation}
For our case, the metric is diagonal, $(x^1, x^2, x^3) \leftrightarrow
(\theta, x, y), N^i \leftrightarrow (N^{\theta}, 0, 0)$ and the metric
is independent of the coordinates $(x, y)$. The $t = $ constant,
hyper-surfaces are diffeomorphic to the 3-torus. The metric components
are given by: $ g_{\theta\theta} = E^x E^y {\cal E}^{-1} = E {\cal
E}^{-2}, g_{xx} = {\cal E}E^y/E^x, g_{yy} = {\cal E}E^x/E^y$ (eq.
(\ref{connectionmetric})).  The Gowdy form of the metric
(\ref{metricmetric}) is realized if one {\em prescribes} $N^{\theta} =
0$ and $N^2 = g_{\theta\theta}$.

Such a prescription is eminently consistent since any metric on a two
dimensional manifold (coordinatized by ($t, \theta$)), can always be
(locally) chosen to be conformally flat. This however does {\em not} fix
the coordinates $t, \theta$ completely -- one can still make the
conformal  diffeomorphisms: $t \to t' = t + \xi^{t}(t, \theta), \theta
\to \theta' = \theta + \xi^{\theta}(t, \theta)$ with $\xi$ satisfying
the conformal Killing equations: $\partial_{t}\xi^t -
\partial_{\theta}\xi^{\theta} = 0 = \partial_{\theta}\xi^t - \partial_t
\xi^{\theta}$.

We will first take the above prescription for the lapse and the shift,
obtain the Hamilton's equations of motion, use the freedom of conformal
diffeomorphisms and reduce the equations to those given in section II.
Subsequently, we will also exhibit gauge fixing functions to arrive at
the same result. This will complete the identification of inequivalent
solutions of the Einstein equation.

With the choices $N^{\theta} = 0, N = \sqrt{E}{\cal E}^{-1}$, the
space-time metric (\ref{SpaceTimeMetric}) is
\begin{equation}
ds^2 ~ = ~ E{\cal E}^{-2}\left( - dt^2 + d\theta^2 \right) + {\cal
E}\left( \f{E^y}{E^x} dx^2 + \f{E^x}{E^y} dy^2 \right)~,
\end{equation}
and the time evolution is governed by the Hamiltonian alone which is
given by,
\begin{eqnarray} 
H[{\cal E}^{-1}\sqrt{E}] & = & \f{1}{\kappa} \int d\theta \left[- \f{1}{
{\cal E} } \bigg\{(K_x E^xK_y E^y)+(K_x E^x + K_y E^y)\cal{E}
\cal{A}\bigg\} \right. \nonumber\\ 
& & \hspace{1.8cm} \left. - \f{1}{4 {\cal E}}\bigg\{
(\del_{\theta}{\cal{E}})^2
-\left({\cal{E}}\del_\theta(\mbox{ln}(E^y/E^x))\right)^2 \bigg\} +
\f{\sqrt{E}}{ {\cal E} } \del_\theta\left(\f{ \cal{E} \del_\theta
\cal{E} }{\sqrt{E}}\right) \right] \label{FixedHam}
\end{eqnarray}

In anticipation let us define $2W := \mathrm{ln}(E^y/E^x)$ and $2a :=
\mathrm{ln}(E^xE^y/{\cal E})$. One obtains,
\begin{equation} 
\f{\dot{E^x}}{E^x} ~ = ~ {\cal E}^{-1}(K_y E^y + {\cal A E}) ~~,~~ 
\f{\dot{E^y}}{E^y} = {\cal E}^{-1}(K_x E^x + {\cal A E}) ~~,~~ 
{\dot{{\cal E}} } = (K_x E^x + K_y E^y) 
\end{equation}
\begin{eqnarray}
2 \partial_t W & := & \partial_t~\mathrm{ln}\f{E^y}{E^x} \hspace{0.8cm}
= ~ \f{( K_x E^x - K_y E^y )}{ {\cal E}} \\
2 \partial_t a & := & \partial_t~\mathrm{ln}\f{E^x E^y}{{\cal E}}
\hspace{0.2cm} = ~ 2 {\cal A} 
\end{eqnarray}

The Poisson brackets of $K_xE^x, K_yE^y$ with the Hamiltonian are given
by,
\begin{eqnarray}
\{K_xE^x, H[{\cal E}^{-1}\sqrt{E}]\} & = &
\f{1}{2}\partial_{\theta}\left({\cal E} \partial_{\theta}
\mathrm{ln}\f{E^y}{E^x}\right) + \f{1}{2} \partial^2_{\theta}{\cal E} \\
\{K_yE^y, H[{\cal E}^{-1}\sqrt{E}]\} & = & -
\f{1}{2}\partial_{\theta}\left({\cal E} \partial_{\theta}
\mathrm{ln}\f{E^y}{E^x}\right) + \f{1}{2} \partial^2_{\theta}{\cal E} \\
\{K_xE^x - K_yE^y, H[{\cal E}^{-1}\sqrt{E}]\} & = &
\partial_{\theta}\left({\cal E} \partial_{\theta}
\mathrm{ln}\f{E^y}{E^x}\right)  \\
\{K_xE^x + K_yE^y, H[{\cal E}^{-1}\sqrt{E}]\} & = &
\partial^2_{\theta}{\cal E}
\end{eqnarray}

From these, we get second order equations for ${\cal E} \ , W$ as,
\begin{eqnarray}
\partial_t^2 {\cal E} & = &  \partial^2_{\theta} {\cal E}
\label{EThetaEqn} \\
\partial_t^2 W & = & \f{1}{ {\cal E}} \partial_{\theta}\left( {\cal E}
\partial_{\theta} W \right) - \left(\f{1}{ {\cal E} } \partial_t {\cal
E}\right) \partial_t W \label{WEqn}
\end{eqnarray}

The equation for ${\cal E}$ is a simple wave equation and given a
solution of this, the equation for $W$ can be solved determining $W$ or
the ratio $E^y/E^x$. From the first order equations, one determines the
$K_xE^x \pm K_yE^y$ as well. The Hamiltonian constraint then determines
${\cal A}$ in terms of known quantities, ${\cal E}, W$ {\em and} the
$\theta$-derivatives of $a$. Using the equation $\partial_t a = {\cal
A}$, one obtains,
\begin{equation}
\partial_t a ~ = ~ {\cal A} = - \f{1}{4}\f{ \partial_t {\cal E}}{ {\cal
E}} + \f{{\cal E}}{\partial_t {\cal E}}\left( (\partial_t W)^2 +
(\partial_{\theta} W)^2 \right) - \f{\partial_{\theta}{\cal
E}}{\partial_t{\cal E}} \partial_{\theta} a  - \f{1}{4} \f{
(\partial_{\theta}{\cal E})^2}{ {\cal E}\partial_t {\cal E}} + \f{
\partial_{\theta}^2 {\cal E}}{\partial_t {\cal E}} \label{aTEqn}
\end{equation}
One can also obtain, by direct computation and using the diffeomorphism
constraint,
\begin{equation}
\partial_{\theta}a ~ = ~ \f{ {\cal E}}{\partial_t {\cal E}}\left[ 2
\partial_t W\partial_{\theta} W - \f{ \partial_{\theta}{\cal E}}{ {\cal
E}} {\partial_t a} + \f{\partial^2_{t\theta}{\cal E}}{ {\cal E}} -
\f{\partial_t{\cal E}\partial_{\theta}{\cal E}}{2 {\cal E}^2} \right]
\label{aThetaEqn}
\end{equation}
From these two equations, one can obtain $\partial_{\theta} a,
\partial_t a$ in terms of ${\cal E}$ and $W$ which can be integrated.
Thus the metric can be completely determined starting from a solution
for ${\cal E}$. 

However, all these solutions are not gauge inequivalent corresponding to
the fact that the coordinates can still be subjected to conformal
diffeomorphisms. Under these coordinate transformations, ${\cal E}$
which is the determinant of the metric on the symmetry torus, is a
scalar. Under conformal diffeomorphisms, the wave operator gets scaled
by a prefactor. Hence, under the transformations generated by conformal
Killing vectors, solutions of the wave equation transform among
themselves. In fact the conformal Killing vectors also satisfy the wave
equation and on the cylinder $(t, \theta)$, both ${\cal E}$ and $\xi$
satisfy the same boundary conditions. Thus, their general solutions are
linear combinations of $exp\{i n (t \pm \theta)\}, n \neq 0$ and a
solution of the form $A + B t$. The Killing vectors however satisfy
first order coupled equations.  This removes the $\theta$-independent,
linear in $t$ piece from the general solution.  Consequently, one can
use conformal diffeomorphisms to remove the $\theta$-dependence from the
solutions for ${\cal E}$ as well as the constant piece. In other words,
all solutions for ${\cal E}$, except ${\cal E} = \# t$ are related to
each other by conformal diffeomorphisms. The gauge inequivalent
solutions are thus obtained from the choice ${\cal E} = t$.
Equivalently, one has finally fixed the ($t, \theta$) coordinates
completely. The time coordinate so fixed will be denoted by $T$. 

With this choice, ${\cal E} = T$, the constraints also simplify to,
\begin{eqnarray}
0 & = & {\cal E} (\partial_{\theta} W)^2 - {\cal E}^{-1} \left\{
K_xE^xK_yE^y + (K_xE^x + K_yE^y){\cal A E} \right\} \\
0 & = & E^x\partial_{\theta}K_x + E^y\partial_{\theta}K_y  \ , 
\end{eqnarray}
one gets $K_xE^x + K_yE^y = 1$ and the equations (\ref{WEqn},
\ref{aTEqn}, \ref{aThetaEqn}) go over to the equations (\ref{Weqn},
\ref{aeqn}). From these one recovers the usual solutions listed in
section II. 

{\em Remark:}
Up to the derivation of the equations for ${\cal E}$ and $W$, the
constraints are not used. The $T^3$ topology has also not been used!
Thus these expressions are also valid for polarized versions of Gowdy
models with other topologies. In Gowdy's original analysis, the three
allowed topologies are distinguished by different choices of solutions
of the equation for ${\cal E}$ ($R$, the determinant of the two metric
on the $T^2$ orbits, in Gowdy's notation).  The different topologies get
distinguished by the boundary conditions on ${\cal E}$ and on the
conformal Killing vectors. For non-$T^3$ topologies, $\theta \in [0,
\pi]$ and ${\cal E}, \xi^{\theta}$ have to vanish at the end-points.
With these taken into account, the gauge inequivalent solutions are
obtained by choosing ${\cal E} = \sin (t)\sin (\theta)$ \cite{Gowdy}.

We reproduced the known results by obtaining the solutions of the
Hamilton's equations with chosen lapse and shift, motivated by
comparison with the space-time form of Gowdy model,  {\em and} invoking
the `residual' freedom in the space-time coordinates to obtain the gauge
inequivalent solutions. Thus we used the canonical structure as well as
the anticipated form of space-time geometry to arrive at the distinct
solutions. We would like to see if the same result can also be derived
by using only the phase space view.

Within a phase space view, the lapse and the shift are to be determined
by doing an explicit gauge fixing. To do this, we will keep the lapse
and the shift as unspecified and look at the evolution generated by the
{\em total} Hamiltonian,
\begin{eqnarray} 
H_{\mathrm{tot}}[N^{\theta}, N] & = & ~ ~  \f{1}{\kappa}\int d \theta
N^{\theta}\left\{ E^x \partial_{\theta} K_x + E^y \partial_{\theta} K_y
- {\cal A} \partial_{\theta}{\cal E} \right\}  \nonumber \\
& & + \f{1}{\kappa} \int d\theta N \left[- \f{1}{ \sqrt{E} } \bigg\{(K_x
E^x K_y E^y)+(K_x E^x + K_y E^y)\cal{E} \cal{A}\bigg\} \right.
\nonumber\\ 
& & \hspace{2.2cm} \left. - \f{1}{4\sqrt{E}}\bigg\{
(\del_{\theta}{\cal{E}})^2
-\left({\cal{E}}\del_\theta(\mbox{ln}(E^y/E^x))\right)^2 \bigg\} +
\del_\theta\left(\f{ \cal{E} \del_\theta \cal{E} }{\sqrt{E}}\right)
\right] \label{UnFixedHam}
\end{eqnarray}

Denoting by over-dots, the Poisson brackets with the total Hamiltonian,
it is straight forward to see,
\begin{eqnarray} 
\f{\dot{E^x}}{E^x} & = & \f{N}{\sqrt{E}}(K_y E^y + {\cal A E}) +
\f{\partial_{\theta}(N^{\theta} E^x)}{E^x} \\
\f{\dot{E^y}}{E^y} & = & \f{N}{\sqrt{E}}(K_x E^x + {\cal A E}) +
\f{\partial_{\theta}(N^{\theta} E^y)}{E^y} \\
\f{\dot{{\cal E}}}{ {\cal E}} & = & \f{N}{\sqrt{E}}(K_x E^x + K_y E^y) +
\f{N^{\theta}\partial_{\theta}{ {\cal E}}}{ {\cal E}} \\
\dot{\left(K_x E^x\right)} & = & \f{1}{2} \partial_{\theta}\left\{ \f{N
{\cal E}}{\sqrt{E}} \left( {\cal E} \partial_{\theta}
\mathrm{ln}\f{E^y}{E^x} + \partial_{\theta} {\cal E} \right) \right\} +
\partial_{\theta}(N^{\theta} K_x E^x) \\
\dot{\left(K_y E^y\right)} & = & \f{1}{2} \partial_{\theta}\left\{ \f{N
{\cal E}}{\sqrt{E}} \left( - {\cal E} \partial_{\theta}
\mathrm{ln}\f{E^y}{E^x} + \partial_{\theta} {\cal E} \right) \right\} +
\partial_{\theta}(N^{\theta} K_y E^y) 
\end{eqnarray}
The following combinations are convenient for looking at gauge fixing.
\begin{eqnarray}
\dot{\left(K_x E^x + K_y E^y\right)} & = & \partial_{\theta}\left\{ \f{N
{\cal E}}{\sqrt{E}} \partial_{\theta} {\cal E} \right\} +
\partial_{\theta}\left\{N^{\theta} \left(K_x E^x + K_y
E^y\right)\right\} \label{ChoiceOne}\\
\dot{{\cal E}} & = & \f{N {\cal E}}{\sqrt{E}}\left(  K_x E^x + K_y E^y
\right) + N^{\theta}\partial_{\theta}{\cal E} \label{ChoiceTwo}\\
\dot{\left(K_x E^x - K_y E^y\right)} & = & \partial_{\theta}\left\{ \f{N
{\cal E}}{\sqrt{E}} {\cal E} \partial_{\theta} \mathrm{ln}\f{E^y}{E^x}
\right\} + \partial_{\theta}\left\{N^{\theta} \left(K_x E^x - K_y
E^y\right)\right\} \label{ChoiceThree}\\
\dot{\left(\mathrm{ln}\f{E^y}{E^x}\right)} & = & \f{N}{\sqrt{E}}\left(
K_x E^x - K_y E^y \right) + N^{\theta}\partial_{\theta}
\mathrm{ln}\f{E^y}{E^x}  \label{ChoiceFour}
\end{eqnarray}

The first two equations above show that we can consistently impose $K_x
E^x + K_y E^y = C_1$, a constant, and $\partial_{\theta}{\cal E} = 0$ as
two gauge fixing conditions. Preservation of the first leads to
$N^{\theta} = f(t)$ while that of the second leads to $N{\cal
E}/\sqrt{E} = g(t)$.  Since $\partial_{\theta}{\cal E} = 0$ already
requires ${\cal E}$ to be a function of $t$ alone, we can strengthen the
gauge fixing condition by specifying ${\cal E} = t$. This determines $N
= C_1\sqrt{E}{\cal E}^{-1}$. Evidently, we must have a non-zero lapse
and therefore $C_1 \neq 0$ must be chosen. The sign of $C_1$ will
determine if ${\cal E}$ increases or decreases with $t$ and by
convention we can take the sign to be positive and without any loss of
generality, we choose $C_1 = +1$ and denote the $t$ by $T$ as before.

The shift is however determined to be a function of $T$ alone. With such
a shift, $C[N^{\theta}] = f(T)\int C$ generates $T$-dependent {\em
translations} of the $\theta$-coordinate. All tensor densities on the
spatial slice, transform as scalars under these translations, and there
is no way to fix the left over constraint $\int C$, by any gauge fixing
condition. However, we can always redefine the $\theta$-coordinate such
that $d\theta - f(t)dt =: d\theta'$. This means that solutions
inequivalent with respect to translations, can be determined by
effectively choosing shift = 0. Incidentally, for other admissible
topologies, the shift has to vanish at $\theta = 0, \pi$ and hence $f(t)
= 0$ is the only admissible solution. We have thus achieved our goal of
determining the {\em same} lapse and shift, by explicit gauge fixing.
The inequivalent solutions are then obtained as in section II.

One can make the physical degrees of freedom explicit by noting that $2W
= \mathrm{ln}(E^y/E^x)$ and $\pi_W := K_xE^x - K_yE^y$ are canonically
conjugate. Similarly, $2\bar{a} := - \mathrm{ln}(E^xE^y)$ and $\pi_{\bar
a} := K_xE^x + K_yE^y$ are also conjugate variables. The gauge fixing
conditions are: ${\cal E} = T, \ \pi_{\bar a} = 1$ while the gauge-fixed
form of constraints become: 
\begin{eqnarray}
C & = & \f{1}{\kappa} \left[ \pi_W \partial_{\theta} W +
\partial_{\theta} \bar a \right] \\
\left(T^{-1} \sqrt{E}\right) H & = & \f{1}{\kappa} \left[ - \f{1 -
\pi_W^2}{4 T} - {\cal A}  + T (\partial_{\theta} W)^2 \right] 
\end{eqnarray}
The Hamiltonian constraint determines ${\cal A}$ completely in terms of
$W, \pi_W$ while the diffeomorphism constraints determines the $\bar a$
{\em except} for the homogeneous ($\theta$-independent) part. The
periodicity of $\bar a$ also requires the $\int \pi_W \partial_{\theta}
W \ = 0$ which is a constraint on the $W, \pi_W$. The physical degrees
of freedom are thus described by $W, \pi_W$ together with one constraint
{\em and} the homogeneous pieces of $\bar a, \pi_{\bar a}$. Our gauge
fixing has fixed the homogeneous part of $\pi_{\bar a}$ to be 1. These
are of course the well known results \cite{Menamarugan}.

Observe that in the homogeneous limit (all variables independent of
$\theta$), one gets the Bianchi I model.  The Hamiltonian constraint,
for each $\theta$ looks like a Bianchi model with a potential and is
highly suggestive of the BKL scenario and has been explored numerically
as well \cite{BKLGowdy}.

This completes the canonical formulation of the polarized Gowdy model on
$T^3$ in terms of the real connection variables.

\section{Discussion} \label{5}

In this paper, two main reformulations of the polarized Gowdy model in
real connection variables have been done. First is the choice of the
gauge invariant variables: $A_x, A_y, E^x, E^y, \alpha, \bar{\alpha}$
and the subsequent canonical transformation to the variables $X, Y,
P^{\xi}, P^{\eta}$. This has already been done in the case of spherical
symmetry and also mentioned for cylindrical waves in \cite{Spherical1}.
The main advantages of these variables are that the volume becomes a
functional of the momenta variables alone and the components of the
connection along the homogeneous directions are separated neatly and
gauge invariantly, into extrinsic curvature components ($X, Y$) and the
spin-connection components ($\Gamma_x, \Gamma_y$). In the quantum
theory, both the features allow a simpler choice of edge and point
holonomies, simpler form for the volume operator and also a more
tractable form of the Hamiltonian constraint \cite{QGowdy}.

The second aspect, obtains the polarized model from the unpolarized by a
simple systematic reduction (Dirac procedure) ensuring a consistent
reduction at the level of {\em physical degrees of freedom}. Getting
this reduction consistently is important since the form of the reduced
constraints, depend on the reducing conditions. In contrast to the
second polarization condition mentioned in the literature, namely
orthogonality of the connection components in analogy with that of the
triad components, our $\chi \approx 0$ condition, (\ref{Constraint2}) is
consistent with dynamics. The consistency is seen in three ways: from a
systematic derivation, verifying the constraint algebra of the reduced
constraints and finally reproducing the known space-times, obtained by
directly solving the Einstein equations for polarized ansatz. We are
thus confident to use these constraint expressions in the passage to
quantization.

We would also like to note that in the reduction to polarized model, we
had two options: $\xi = 0 \ (E^{\theta}_3 > 0)$ or $\xi = \pi \
(E^{\theta}_3 < 0)$.  In the metric variables and classically, either
one of these suffices.  (In the triad variables, these two correspond to
opposite orientations.) The subsequent gauge fixing was also naturally
restricted to one of these choices (we chose the former). At this stage,
one could imagine doing a ``loop quantization'' of the gauge-fixed model
which now has a true Hamiltonian and explore the fate of the
singularity.  In a quantum theory however, one could have an extension
across the degenerate triad and this will be missed in a quantization of
the gauge-fixed model. 

\begin{acknowledgments} 
Discussions with Alok Laddha and Martin Bojowald are gratefully
acknowledged.
\end{acknowledgments}

\appendix*
\section{}

In this appendix, we collect some of the useful expressions to help
reproduce the computations in the main text.

The symmetry reduction leaves only the following non-zero components of
the gravitational connection and the densitized triad:
\begin{eqnarray}
A^i_a ~ \longrightarrow ~ A^3_{\theta}, A^I_{\rho}, ~ ~ I = 1, 2 ~ ~
\rho = x, y \ , \nonumber \\
E_i^a ~ \longrightarrow ~ E_3^{\theta}, E_I^{\rho}, ~ ~ I = 1, 2 ~ ~
\rho = x, y \ .
\end{eqnarray}
The triad has the components: $e^a_i = E^a_i E^{-1/2}$ with $E :=
\mathrm{det}E^a_i = E^{\theta}_3 \Delta, \ \Delta := E^x_1 E^y_2 - E^x_2
E^y_1$. The co-triad (inverse triad), $e_a^i$, has the components:
$e^3_{\theta} = \sqrt{\Delta/E^{\theta}_3}, \ e^1_x =
\sqrt{E^{\theta}_3/\Delta} \ E^y_2, \ e^1_y = -
\sqrt{E^{\theta}_3/\Delta} \ E^x_2, \ e^2_x = -
\sqrt{E^{\theta}_3/\Delta}  \ E^y_1, \ e^2_y =
\sqrt{E^{\theta}_3/\Delta}  \ E^x_1$.

The spin connection is defined by,
\begin{equation} \label{SpinConnection}
\Gamma^i_a ~ := ~ - \epsilon^{ijk} e^b_j \left( \partial_{[a} e^k_{b]} +
\f{1}{2} e^c_k e^l_a \partial_{[c} e^l_{b]} \right) \ .
\end{equation}
Of these, $\Gamma^3_{\rho} = 0 = \Gamma^I_{\theta}$ are identically
zero. The remaining components are given by,
\begin{eqnarray}
\Gamma^3_{\theta} & = & \f{1}{2 \Delta}\left( 
E^x_1 \partial_{\theta} E^y_1  - 
E^y_1 \partial_{\theta} E^x_1  + 
E^x_2 \partial_{\theta} E^y_2  - 
E^y_2 \partial_{\theta} E^x_2 \right) \nonumber \\
\Gamma^1_x & = & \f{1}{2}\sqrt{\f{E^{\theta}_3}{\Delta}}\left[
\partial_{\theta} \left(\sqrt{\f{E^{\theta}_3}{\Delta}} E^y_1\right) +
\f{E^{\theta}_3}{\Delta} \vec{E^y}\cdot\vec{E^y} \partial_{\theta}
\left(\f{E^x_2}{\sqrt{E}}\right) - 
\f{E^{\theta}_3}{\Delta} \vec{E^x}\cdot\vec{E^y} \partial_{\theta}
\left(\f{E^y_2}{\sqrt{E}}\right) \right] \nonumber \\
\Gamma^2_x & = & \f{1}{2}\sqrt{\f{E^{\theta}_3}{\Delta}}\left[
\partial_{\theta} \left(\sqrt{\f{E^{\theta}_3}{\Delta}} E^y_2\right) -
\f{E^{\theta}_3}{\Delta} \vec{E^y}\cdot\vec{E^y} \partial_{\theta}
\left(\f{E^x_1}{\sqrt{E}}\right) + 
\f{E^{\theta}_3}{\Delta} \vec{E^x}\cdot\vec{E^y} \partial_{\theta}
\left(\f{E^y_1}{\sqrt{E}}\right) \right] \\
\Gamma^1_y & = & \f{1}{2}\sqrt{\f{E^{\theta}_3}{\Delta}}\left[
- \partial_{\theta} \left(\sqrt{\f{E^{\theta}_3}{\Delta}} E^x_1\right) -
\f{E^{\theta}_3}{\Delta} \vec{E^x}\cdot\vec{E^y} \partial_{\theta}
\left(\f{E^x_2}{\sqrt{E}}\right) + 
\f{E^{\theta}_3}{\Delta} \vec{E^x}\cdot\vec{E^x} \partial_{\theta}
\left(\f{E^y_2}{\sqrt{E}}\right) \right] \nonumber \\
\Gamma^2_y & = & \f{1}{2}\sqrt{\f{E^{\theta}_3}{\Delta}}\left[
- \partial_{\theta} \left(\sqrt{\f{E^{\theta}_3}{\Delta}} E^x_2\right) +
\f{E^{\theta}_3}{\Delta} \vec{E^x}\cdot\vec{E^y} \partial_{\theta}
\left(\f{E^x_1}{\sqrt{E}}\right) - 
\f{E^{\theta}_3}{\Delta} \vec{E^x}\cdot\vec{E^x} \partial_{\theta}
\left(\f{E^y_1}{\sqrt{E}}\right) \right] \nonumber 
\end{eqnarray}
where, $\vec{E^x}\cdot\vec{E^y} := E^x_1 E^y_1 + E^x_2 E^y_2$ etc.

In terms of the radial and angular variables $E^x, E^y, E^{\theta}_3 (=
{\cal E}), \xi, \eta$ given in equations (\ref{AngleOne},
\ref{AngleTwo}, \ref{CanOne}), one has $\Delta = E^x E^y
\mathrm{cos}\xi, \ \vec{E^x}\cdot\vec{E^x} = (E^x)^2,
\vec{E^y}\cdot\vec{E^y} = (E^y)^2, \ \vec{E^x}\cdot\vec{E^y} = E^x E^y
\mathrm{sin}\xi$. 

In the computation of the Hamiltonian constraint, one needs the
combinations: $E^a_i \Gamma^i_b$ and $E^a_i A^i_b$. The non-zero ones
are given by,
\begin{eqnarray}
E^{\theta}_i \Gamma^i_{\theta} & = & \f{1}{2}E^{\theta}_3 \left[
\mathrm{tan} \xi \ \partial_{\theta}\left(\mathrm{ln}\f{E^y}{E^x}\right)
- \partial_{\theta} \eta \right] \nonumber \\
E^x_i \Gamma^i_x & = & \f{1}{2}\partial_{\theta}\left( E^{\theta}_3
\mathrm{tan} \xi \right) \hspace{0.95cm} , ~ ~ 
E^y_i \Gamma^i_y ~ = ~ - \f{1}{2}\partial_{\theta}\left( E^{\theta}_3
\mathrm{tan} \xi \right) \nonumber \\
E^x_i \Gamma^i_y & = & - \f{1}{2}\partial_{\theta}\left(
\f{E^x}{E^y}\f{E^{\theta}_3}{\mathrm{cos} \xi} \right) ~ ~ , ~ ~
E^y_i \Gamma^i_x ~ = ~ \f{1}{2}\partial_{\theta}\left(
\f{E^y}{E^x}\f{E^{\theta}_3}{\mathrm{cos} \xi} \right) \\
E^x_i A^i_x & = & K_x E^x \hspace{0.95cm} , ~ ~ E^y_i A^i_y ~ = ~ K_y E^y
\hspace{0.95cm} , ~ ~ E^{\theta}_i A^i_{\theta} ~ = ~ E^{\theta}_3
A^3_{\theta} \ , \nonumber \\
E^x_i A^i_y & = & \left(\f{E^y}{E^x}\right)\left\{ - (P^{\xi} +
P^{\eta}) \mathrm{cos} \xi + K_x E^x \mathrm{sin} \xi \right\} \nonumber
\\ E^y_i A^i_x & = & \left(\f{E^x}{E^y}\right)\left\{ (P^{\eta} -
P^{\xi}) \mathrm{cos} \xi + K_y E^y \mathrm{sin} \xi  \right\}
\end{eqnarray}

The Hamiltonian constraint in eqn (\ref{hamiltonian1}) is simplified by
eliminating the extrinsic curvature in terms of the gravitational
connection and the spin-connection, $K_a^i := \gamma^{-1} ( A^i_a -
\Gamma^i_a)$, and using the above equations.  One begins with the
expression:
\begin{eqnarray}
H & = & \f{1}{2\kappa}\f{1}{\sqrt{E}}\left[ ~
\epsilon_{ijk}E^a_iE^b_j(\partial_aA^k_b - \partial_bA^k_a) +
\epsilon_{ijk}\epsilon^k_{\ jk}E^a_iE^b_jA^m_aA^n_b  \right.
\label{HamOriginal} \\
& & \hspace{1.5cm}\left. - (1 + \gamma^{-2})\left\{ E^a_i(A^i_a -
\Gamma^i_a)E^b_j(A^j_b - \Gamma^j_b)  - E^a_i(A^j_a -
\Gamma^j_a)E^b_j(A^i_b - \Gamma^i_b) \right\} \right] \nonumber\\
& = & \f{1}{2\kappa}\f{1}{\sqrt{E}}\left[ ~
\left\{\epsilon_{ijk}E^a_iE^b_j(\partial_aA^k_b - \partial_bA^k_a)
\right\}\right. \label{HamExpanded} \\
& & \hspace{1.5cm}- \gamma^{-2}\left\{ (E^a_iA^i_a)(E^b_jA^j_b) -
(E^a_iA^i_b)(E^b_jA^j_a)\right\} \nonumber \\
& & \hspace{1.5cm}- (1 + \gamma^{-2})\left\{
(E^a_i\Gamma^i_a)(E^b_j\Gamma^j_b) -
(E^a_i\Gamma^i_b)(E^b_j\Gamma^j_a)\right\} \nonumber \\
& & \hspace{1.5cm}\left.- 2(1 + \gamma^{-2})\left\{ -
(E^a_iA^i_a)(E^b_j\Gamma^j_b) + (E^a_i\Gamma^i_b)(E^b_jA^j_a)\right\}
\right] \nonumber 
\end{eqnarray}
The terms quadratic in $A$'s combine to get $\gamma^{-2}$, while terms
linear in $A$ and $\Gamma$ get a prefactor of $2(1 + \gamma^{-2})$.  In
terms of the angular and radial variables given in (\ref{AngleOne}), the
terms in the braces in eqn (\ref{HamExpanded}) become:
\begin{eqnarray}
\mathrm{First} & = & 2 \epsilon_{JK} E^{\theta}_3 E^{\rho}_J
\partial_{\theta} A^K_{\rho} \nonumber\\
& = & E^{\theta}_3 \ \left[\ (K_x E^x + K_y E^y)\partial_{\theta}\eta +
(K_xE^x - K_y E^y) \partial_{\theta}\xi - 4 \partial_{\theta}P^{\eta}
\right. \nonumber \\ 
& & \hspace{0.8cm}\left.  + 2
P^{\eta}\partial_{\theta}\mathrm{ln}(E^xE^y) - 2
P^{\xi}\partial_{\theta}\mathrm{ln}(E^y/E^x) \right] \\
\mathrm{Second}  & = & \left(E^{\theta}_3A^3_{\theta} + K_xE^x +
K_yE^y\right)^2 - \left(E^{\theta}_3A^3_{\theta}\right)^2 -
(E^{\rho}_IA^I_{\sigma})(E^{\sigma}_JA^J_{\rho}) \nonumber \\
& = & 2\  \mathrm{cos}^2(\xi) \left(K_x E^x K_y E^y + (P^{\eta})^2 -
(P^{\xi})^2 \right) ~ + ~ 2 \ E^{\theta}_3A^3_{\theta}\left(K_xE^x +
K_yE^y\right)  \nonumber\\ & & \hspace{0.0cm} + ~ 2\ \mathrm{sin}(\xi)
\mathrm{cos}(\xi) \left(P^{\xi}(K_xE^x + K_yE^y) - P^{\eta}(K_xE^x -
K_yE^y) \right) \\
\mathrm{Third} & = & -
\left(E^{\rho}_I\Gamma^I_{\sigma}\right)\left(E^{\sigma}_J\Gamma^J_{\rho}\right)
\nonumber \\
& = & \f{1}{2}\left[ \left(\partial_{\theta}E^{\theta}_3\right)^2
- \left(\f{E^{\theta}_3} {\mathrm{cos}(\xi)}
\partial_{\theta}(\xi)\right)^2
- \left(\f{E^{\theta}_3} {\mathrm{cos}(\xi)}
\partial_{\theta}\mathrm{ln}(E^y/E^x)\right)^2\right] \\
\mathrm{Fourth} & = & -
(E^{\theta}_3\Gamma^3_{\theta})(E^{\theta}_3A^3_{\theta} + K_xE^x +
K_yE^y) + (E^{\theta}_3\Gamma^3_{\theta})(E^{\theta}_3A^3_{\theta}) +
(E^{\rho}_I \Gamma^I_{\sigma})(E^{\sigma}_JA^J_{\rho}) \nonumber \\
& = & \f{E^{\theta}_3}{2}\left[ - 2 P^{\xi} \
\partial_{\theta}\mathrm{ln}(E^y/E^x)\  +\  2 P^{\eta} \
\partial_{\theta}\mathrm{ln}E^{\theta}_3\  +\  2
P^{\eta}\mathrm{tan}(\xi) \partial_{\theta}\xi \right. \nonumber \\ & &
\hspace{0.8cm} \left.  + \ (K_xE^x + K_yE^y) \partial_{\theta}\eta \ + \
(K_xE^x - K_yE^y)\partial_{\theta}\xi \right]
\end{eqnarray}
Using these equations and with further rearrangement of terms, leads
finally to the Hamiltonian constraint of eqn. (\ref{hamiltonian2}).

The polarization conditions are $\xi = 0 = 2 P^{\xi} +
E^{\theta}_3\partial_{\theta} \mathrm{ln}(E^y/E^x)$. With $\xi = \beta -
\bar{\beta} = 0$ alone, the spin-connection components simplify as:
$\Gamma^3_{\theta} = -(1/2)\partial_{\theta} \eta$, 
\begin{eqnarray}
\Gamma^1_x ~ = ~ - \mathrm{sin}\beta\ \Gamma_x ~ ~,~ ~ 
\Gamma^2_x ~ = ~  \mathrm{cos}\beta\ \Gamma_x  & ~;~ & 
\Gamma^1_y ~ = ~  \mathrm{cos}\beta\ \Gamma_y ~ ~,~ ~ 
\Gamma^2_y ~ = ~  \mathrm{sin}\beta\ \Gamma_y  \label{GammaComp1} \\
\Gamma_x ~ := ~ \f{1}{2}\f{E^{\theta}_3}{E^x}
\partial_{\theta}\mathrm{ln}\left(E^{\theta}_3 \f{E^y}{E^x}\right)
\hspace{1.7cm} & ~~~,~~~ &
\Gamma_y ~ := ~ \f{1}{2}\f{E^{\theta}_3}{E^y}
\partial_{\theta}\mathrm{ln}\left(\f{1}{E^{\theta}_3}
\f{E^y}{E^x}\right) \label{GammaComp2} 
\end{eqnarray}
Using the Gauss constraint and the polarization constraint, $\chi = 0$,
it follows that,
\begin{equation} \label{AComp}
A_x \mathrm{sin}\ \alpha := - \f{P^{\beta}}{E^x} =  \Gamma_x ~ ~ , ~ ~
A_y \mathrm{sin}\ \bar{\alpha} := - \f{P^{\bar{\beta}}}{E^y} =  -
\Gamma_y \ .
\end{equation} Notice that $\Gamma_x, \Gamma_y$ are {\em gauge
invariant}.

In checking preservation of various constraints as well as verifying
the constraint algebra, the following Poisson bracket is useful,
\begin{eqnarray}
\left\{K_xE^x(\theta), \f{\partial_{\theta '}
E^x(\theta')}{E^x(\theta')}\right\} & = &
\kappa\partial_{\theta'}\left(\f{E^x(\theta)}{E^x(\theta')}
\delta(\theta - \theta')\right) ~ ~ , ~ ~ \kappa := \f{2
G_{\mathrm{Newton}}}{\pi} \ .
\end{eqnarray}


\begin{thebibliography}{10}

\bibitem{Gowdy}
%
Gowdy R H, 1974 Vacuum space-times with two parameter space-like
isometry groups and compact invariant hypersurfaces: Topologies and
boundary conditions, {\em Ann. Phys.} {\bf 83} 203-241.

\bibitem{Moncrief}
%
Moncrief V, 1981 Infinite-dimensional family of vacuum
cosmological models with Taub-NUT (Newman-Unti-Tamburino)-type
extensions {\em Phys. Rev} {\bf D 23} 312-315.

\bibitem{IsenbergMoncrief}
%
Isenberg J and Moncrief V, 1990 Asymptotic behavior of the
gravitational field and the nature of singularities in Gowdy space-time
{\em Ann. Phys.} {\bf 199} 84-122.

\bibitem{Berger}
%
Berger, B K, 2002, Asymptotic Behavior of a Class of Expanding Gowdy
Spacetimes, [gr-qc/0207035]. 

\bibitem{ADMQuantization}
Misner C W,  1973 A Minisuperspace Example: The Gowdy $T^{3}$ Cosmology
{\em Phys. Rev.} {\bf D 8} 3271-3285;
\\%
Berger B K, 1975 Quantum cosmology: Exact solution for the Gowdy $T^{3}$ model
{\em Phys. Rev.} {\bf D 11} 2770-2780.
%

\bibitem{Pierri}
Pierri M,  2002  Probing quantum general relativity through exactly 
soluble midi-superspaces. II: Polarized Gowdy models
{\em Int. J. Mod. Phys.} {\bf D11} 135,
[gr-qc/0101013].
%

\bibitem{MenamaruganTorre}
Corichi A, Cortez J and Quevedo H, 2002
On unitary time evolution in Gowdy $T^3$ cosmologies
{\em Int. J. Mod. Phys.} {\bf D 11} 1451-1468
[gr-qc/0204053];
\\%
Torre C G, 2002  Quantum dynamics of the polarized Gowdy $T^3$ model
{\em Phys. Rev.} {\bf D 66} 084017
[gr-qc/0206083];
\\%
Torre C G, 2006  Observables for the polarized Gowdy model
{\em Class. Quant. Grav.} {\bf 23} 1543-1556
[gr-qc/0508008];
\\%
Torre C G, 2007  Schroedinger representation for the polarized Gowdy model
{\em Class. Quant. Grav.} {\bf 24} 1-13 [gr-qc/0607084];
\\%
Cortez J and Mena Marugan G A,  2005
Feasibility of a unitary quantum dynamics in the Gowdy $T^3$ cosmological model
{\em Phys. Rev.} {\bf D 72} 064020
[gr-qc/0507139].
%

\bibitem{Corichi}
Corichi A, Cortez J and Mena Marugan G A,  2006
Unitary evolution in Gowdy cosmology
{\em Phys. Rev.} {\bf D 73} 041502
[gr-qc/0510109];
\\%
Corichi A, Cortez J and Mena Marugan G A,  2006
Quantum Gowdy $T^3$ model: A unitary description
{\em Phys. Rev.} {\bf D 73}  084020
[gr-qc/0603006];
\\%
Corichi A, Cortez J, Mena Marugan G A and Velhinho J M,  2006
Quantum Gowdy $T^3$ model: A uniqueness result
{\em Class. Quant. Grav.}  {\bf 23} 6301
[gr-qc/0607136];
\\%
Cortez J, Mena Marugan G A and Velhinho J M,  2007
Uniqueness of the Fock quantization of the Gowdy $T^3$ model
{\em Phys. Rev.} {\bf D 75} 084027
[gr-qc/0702117];
\\%
Corichi A, Cortez J, Mena Marugan G A and Velhinho J M,  2007
Quantum Gowdy $T^3$ Model: Schrodinger Representation with Unitary Dynamics
{\em Phys. Rev.}  {\bf D 76} 124031
[arXiv:0710.0277].

\bibitem{HusainSmolin}
%
Husain V and Smolin L, 1989 Exactly Soluble Quantum Cosmologies
from Two Killing Field Reductions of General Relativity {\em Nucl.
Phys.} {\bf B 327} 205-238.

\bibitem{Husain}
%
Husain V, 1994, Observables for spacetimes with two Killing field
symmetries, {\em Phys. Rev.} {\bf D 50}, 6207-6216.

\bibitem{Menamarugan}
%
Mena Marugan G A, 1997 Canonical quantization of the Gowdy model
{\em Phys. Rev.} {\bf D 56} 908-919, [gr-qc/9704041].

\bibitem{HusainSingularity}
%
Husain V, 1987, Quantum effects on the singularity of the Gowdy
cosmology {\em Class. Quantum Grav.} {\bf 4}, 1587-1591.

\bibitem{BergerGarfinkle}
%
Berger B K and Garfinkle D, 1998 Phenomenology of the Gowdy
Universe on $T^3 \times R$ {\em Phys. Rev.} {\bf D 57} 4767-4777,
[gr-qc/9710102].

\bibitem{AshtekarLewandowski} Ashtekar A and Lewandowski J, 2004,
Background Independent Quantum Gravity: A Status Report, {\em Class.
Quant. Grav.}, {\bf 21}, R53, [gr-qc/0404018].

\bibitem{Spherical1}
Bojowald M, 2004
Spherically symmetric quantum geometry: States and basic operators,
{\em Class. Quant. Grav.}, {\bf 21}, 3733-3753,
[gr-qc/0407017].
%

\bibitem{Neville}
Neville D E, 2006 
The volume operator for singly polarized gravity waves with planar
or cylindrical symmetry, {\em Phys. Rev}, {\bf D 73}, 124005, 
[gr-qc/0511006]. 

\bibitem{BKLGowdy} 
%
Berger, B K, 2002, Numerical Approaches to Spacetime Singularities {\em
Living Rev. Relativity}, {\bf 5}, 1, [gr-qc/0201056].

\bibitem{QGowdy}
Banerjee, K and Date, G, 2007, Loop Quantization of Polarized Gowdy
Model on $T^3$: Quantum Theory, [arXiv:0712.0687].

\end{thebibliography}
\end{document}